\documentclass[useAMS,usenatbib,twocolumn]{mn2e}

\usepackage[latin1]{inputenc}  
\usepackage[T1]{fontenc}       
\usepackage[dvips]{graphicx}
\usepackage{amsmath}
\usepackage{amssymb}
\usepackage{indentfirst}

\title[The steep decay phase of GRB080503]{The long rapid decay phase of the extended emission from the short GRB\,080503.}

\author[F. Genet, N.R. Butler and J.Granot]{F. Genet$^{1,2}$\thanks{E-mail:
f.genet@herts.ac.uk;}, N.R. Butler$^{3}$ and J.Granot$^{1}$\\
$^{1}$ Center for Astrophysics Research, University of Hertfordshire, UK.\\
$^{2}$ Racah institute of Physics, Hebrew University of Jerusalem, Israel.\\
$^{3}$ Astronomy Department, University of California, Berkeley, CA, USA}
\begin{document}


\pagerange{\pageref{firstpage}--\pageref{lastpage}} \pubyear{2009}

\maketitle

\label{firstpage}

\begin{abstract}
The gamma-ray burst (GRB) 080503 was classified as a short GRB with extended
emission (Perley et al. 2009). The origin of such extended emission (found in
about a quarter of Swift short GRBs) is still unclear and may provide some
clues to the identity of the elusive progenitors of short GRBs. The extended
emission from GRB\,080503 is followed by a rapid decay phase (RDP) that is
detected over an unusually large dynamical range (one decade in time and $\sim
3.5$ decades in flux), making it ideal for studying the nature of the extended
emission from short GRBs. We model the broad envelope of extended emission and
the subsequent RDP using a physical model for the prompt GRB emission and its
high latitude emission tail (Genet \& Granot 2009), in which the prompt
emission (and its tail) is the sum of its individual pulses (and their
tails). For GRB\,080503, a single pulse fit is found to be unacceptable, even
when ignoring short timescale variability.  The RDP displays very strong
spectral evolution and shows some evidence for the presence of two spectral
components with different temporal behaviour, likely arising from distinct
physical regions. A two pulse fit (a first pulse accounting for the gamma-ray
extended emission and decay phase, and the second pulse accounting mostly for
the X-ray decay phase) provides a much better (though not perfect) fit to the
data. The shallow gamma-ray and steep hard X-ray decays are hard to account
for simultaneously, and require the second pulse to deviate from the simplest
version of the model we use. Therefore, while high latitude emission is a
viable explanation for the RDP in GRB\,080503, it does not pass our tests with
flying colors, and it is quite plausible that another mechanism is at work
here. Finally, we note that the properties of the RDP following the extended
emission of short GRBs (keeping in mind the very small number of well studied
cases so far) appear to have different properties than that following the
prompt emission of long GRBs. However, a larger sample of short GRBs with
extended emission is required before any strong conclusion can be drawn.
\end{abstract}


\section{Introduction}

GRB080503 was detected by the {\it Swift} Burst Alert Telescope (BAT) on 2008
may 3 \citep{peretal09}. Its prompt gamma-ray emission presents a short ($\sim
0.32\;$s) intense initial spike followed by an extended emission lasting
several minutes, for a total duration of $T_{\rm 90} \approx 232\;$s. The
first short spike duration in the $15-150\;$keV band is $0.32 \pm 0.07\;$s and
its peak flux is $(1.2\pm0.2)\times 10^{-7}\;$erg$\;$cm$^{-2}\;$s$^{-1}$. The
count rate hardness ratio between the $50-100\;$keV and the $25-50\;$keV of
the initial spike is $1.2\pm0.3$, consistent with other short {\it Swift}
bursts, but also with some long bursts. The fluence of the extended emission
(measured between $5\;$s and $140\;$s) in the $15-150\;$keV band is
$(1.86\pm0.14)\times 10^{-6}\;$erg$\;$cm$^{-2}$, about thirty times that of
the initial spike, higher than for any other short {\it Swift} bursts, but
within the range of such ratio measured for BATSE short
bursts. \citet{peretal09} found a spectral lag between the $50-100\;$keV and
the $25-50\;$keV bands consistent with zero. All this led them to associate
GRB080503 with the ``short'' \citep{kouvetal93} class.

The {\it Swift} X-Ray Telescope (XRT) started observing GRB080503 about
$82\;$s after the burst, detecting a bright early afterglow that rapidly
decayed ($\alpha=2 - 4$ with $F_{\nu} \propto t^{-\alpha}$) to below the
detection threshold during the first orbit, which makes a record overall
decline of $\sim 6.5$ decades, with a steep decay clearly observed for $\sim
3.5$ decades (see figure 6 of Perley et al. 2009). Chandra detections at $\sim
3\;$days after the GRB indicate the presence of a separate X-ray afterglow
component $\sim 10^6$ times fainter than the peak emission from the X-ray tail
of the prompt extended emission. The gamma-ray lightcurve of the extended
emission (which can be found with the X-ray light-curve of the extended
  emission in figure \ref{fig_fits}) presents two relatively well defined
spikes at $\sim 28\;$s and $\sim 36\;$s, followed by some less defined
variability with several narrow spikes (see also figure 1 of Perley et
al. 2009). The X-ray RDP lightcurves present a first break at $\sim 180\;$s
seen in both soft and hard bands; later on, the hard X-ray band clearly shows
a steepening (to a temporal slope of about $-4.2$) that is only marginally
seen in the soft X-rays. The simultaneous gamma-ray decay is much shallower,
with a temporal slope only about $-2$.

The rapid decay phase (RDP) of GRB080503 is a smooth temporal and spectral
continuation of the extended emission in the gamma-rays (Perley et al.,
2009). This suggests that it is the tail of the prompt emission (O'Brien et
al., 2006; Butler \& Kocevski, 2007). The most popular model to explain the
RDP following the prompt emission of long GRBs is High Latitude Emission (HLE;
Kumar and Panaitescu 2000). In this model, after the prompt emission stops, photons
from increasingly larger angles relative to the line of sight still reach the
observer, due to the curvature of the (assumed to be quasi-spherical) emitting
surface. The blueshift of such photons decreases with the angle from the
line of sight, and therefore with their arrival time. This results in a simple
relation between the temporal and spectral indexes of the flux, $\alpha = 2+
\beta$, where $F_{\nu} \propto t^{-\alpha} \nu^{-\beta}$, that holds at late
times (when $t-t_0 \gg \Delta t$, where $t_0$ is the onset time of the pulse
and $\Delta t$ its width) for each pulse of the prompt emission. 

The RDP following the extended emission of the short GRB\,080503, however,
shows strong spectral evolution that is different than that typically seen in
the RDP following the prompt emission of long GRBs. Whereas in the latter case the RDP
usually shows a monotonic softening of the spectrum with time (see Zhang, Liang \& Zhang
2007 and references therein), the RDP in GRB\,080503 shows a similar softening
in the XRT energy range together with a hardening in the spectral slope
between the XRT and BAT energy ranges. This behavior suggests the presence of
two distinct spectral components. 

Such extended emission is observed in about a quarter of Swift short bursts
(Norris and Gehrels 2009; Norris et al. 2009). It usually lasts tens of seconds, and has
been shown to be always softer than the initial short spike, however showing
negligible lags as the initial spike \citep{nobo06}. The same
authors also found that the extended emission is softer than the initial
spike, making it closer to long bursts, which are softer than short
bursts. However, the initial spike of short burst with and without EE show
similar properties \citep{noge09}. Many authors recently turned to
environmental constraints to explain differences between short GRBs with and
without EE, but with results at best uncertain, sometime contradictory (Rhoads
2008; Troja et al. 2008; Fong, Berger \& Fox 2009; Sakamoto \& Gehrels 2009;
Nysewander, Fruchter \& Pe'er 2009). There is as of now no strong explanation
for the mechanism behind extended emission.  It is therefore very important to
test whether HLE can indeed explain the RDP of GRB080503, as this would allow
some comparison with results for the RDP following the prompt emission without
extended emission (Willingale et al., 2009). This will stress similarities and
differences between the extended emission and the prompt emission itself,
ultimately leading to a better understanding of the origin of extended
emission.

Genet \& Granot (2009; hereafter GG09) have developed a simple yet physical
and self-consistent model for the prompt and high latitude emission. The very
large dynamical range over which the RDP of GRB080503 is observed allows us to
test its behaviour at late times, which is rarely possible in other {\it
  Swift} bursts. In order to test whether the extended emission is coming from
a mechanism similar to the prompt emission itself, we apply the GG09 model to
GRB080503 data by folding it through the response matrices generated from the
four energy bands of data (0.3-1.3 keV and 1.3-10.0 keV for the XRT, 15-50 keV
and 50-150 keV for the BAT). Section \S$\;$\ref{sec_modeldescription}
summarizes the theoretical model used to fit the burst lightcurves. Section
\S$\;$\ref{sec_datafit} describes how the data are reduced and fitted and the
results of our fits, and in section \S$\;$\ref{sec_discussion} we discuss the
implications of these results.

\section{The theoretical model} \label{sec_modeldescription}

The prompt and extended emissions and their tail (the RDP) are taken to be the
sum of its pulses and their tails, as described in GG09. Each pulse represents
a single emission episode, assumed to come from an ultra-relativistic ($\gamma
\gg 1$) thin spherical expanding shell, turns on at radius $R_0$ and turns
abruptly off at radius $R_f \equiv R_0 + \Delta R$ where $\Delta R$ is the
width of the emission region. The emission is assumed to be uniform over the
emitting shell and isotropic in its rest frame. The observed flux is
calculated by integrating over the surface of equal arrival time of photons to
the observer (following Granot 2005 and Granot et al. 2008). The
characteristic observed times of a pulse are the ejection time of the emitting
shell ($T_{\rm ej}$), the radial time at $R_0$ ($T_0$) and at $R_f$
($T_f\equiv T_0 (1+\Delta R/R_0)$); $T_f$ is also essentially the temporal
width of the pulse. For $\Delta R/R_0 \lesssim$ a few, the peak of the pulse
is observed at $T_{\rm peak} = T_{\rm ej}+T_f$. We consider that the peak
luminosity of the shell evolves as $L'_{\nu'_p} \propto R^{a}$. In the case of
emission by synchrotron mechanism from electrons accelerated in internal
shocks and cooling fast, $a=1$, but in the following it will be kept free as
fixing its value does not simplify calculations and it allows to check for
deviations from these assumptions. In this framework we also have the peak
frequency of the $\nu F_{\nu}$ spectrum evolving as $\nu'_p \propto
R^{-1}$. The emission spectrum is assumed to be the phenomenological Band
function \citep{bandetal93}. The number of photons $N$ per unit photon energy
$E$, area $A$ and observed normalized time $\tilde{T} \equiv (T-T_{\rm
  ej})/T_0$ of a single emission episode is then (where we also define
$\tilde{T}_f \equiv (T_f-T_{\rm ej})/T_0$):
\begin{eqnarray*} \label{eq_dN/dtdEdA}
\frac{d N}{dE dA dT}(E,\tilde{T}\geq 1) =
\tilde{T}^{-1}
\left[\min \left(\tilde{T},\tilde{T}_f\right)^{a+2}-1\right]
B\left(\frac{E}{E_0}\tilde{T}\right),
\end{eqnarray*}
where $a$ is kept free to vary as fixing its value does not simplify calculations and
\begin{eqnarray*} \label{eq_bandfunction_energy}
B(z) = B_{\rm norm} \left\{ \begin{array}{ll}
z^{-1-\beta_l} e^{-z}  &  z \leqslant \Delta \beta\,\\
z^{-1-\beta_h} (\Delta \beta)^{\Delta \beta} e^{-\Delta \beta}  &  z \geqslant \Delta \beta\,
\end{array} \right.
\end{eqnarray*}
is the Band function with a normalization constant $B_{\rm norm}$, with high and
low energy photon indexes of $-1-\beta_h$ and $-1-\beta_l$ respectively, where
$\Delta \beta=\beta_h-\beta_l$ and $z = (E/E_0)\tilde{T}$. One can remark
that the observed spectrum is a Band function spectrum, as the emitted one,
but with its peak energy sweeping through the observed bands: the peak of
$E^2dN/dE$ is $E_{\rm peak} = (1-\beta_l) E_0/\tilde{T}$ (where $E_0 \equiv
E_{\rm peak}(\tilde{T}=1)/(1-\beta_l)$), which decreases linearly with the
normalized time $\tilde{T}$. At times $\tilde{T} \geqslant \tilde{T}_f$ is
observed the high latitude emission of the shell. In this part of the
lightcurve of the pulse, the spectral slope $-d\log (E^2 dN/dtdEdA)/d\log E$
will evolve from $1-\beta_l-\tilde{T}E_{\rm obs}/E_0$ (where $E_{\rm obs}$ is
the observed photon energy) to $1-\beta_h$ if $E_{\rm obs}<E_0$. This creates
a break in the lightcurve, which becomes very smooth if the observations are
done in an energy band of finite width.

\section{Data Reduction and Fitting} \label{sec_datafit}

We download the raw, unfiltered Swift BAT and XRT data for GRB~080503 from the
{\it Swift}~Archive\footnote{ftp://legacy.gsfc.nasa.gov/swift/data}. Our
reduction of these data to science quality light curves and spectra are
detailed in \citet{peretal09}. Here, we subdivide the X-ray and $\gamma$-ray
light curves into much finer time bins to study the spectral evolution. We
begin with the BAT 15-350 keV and XRT 0.3-10.0 keV lightcurves, with each bin
containing sufficient counts to reach a signal-to-noise ratio ($S/N$) of 3 or
greater. We then generate response matrices for each time bin, using the tools
summarized in \cite{peretal09}. Our reduction accounts for the spacecraft slew
for BAT as well as photon pileup for the XRT, among other effects.

\begin{figure}
\includegraphics[width=3.5in]{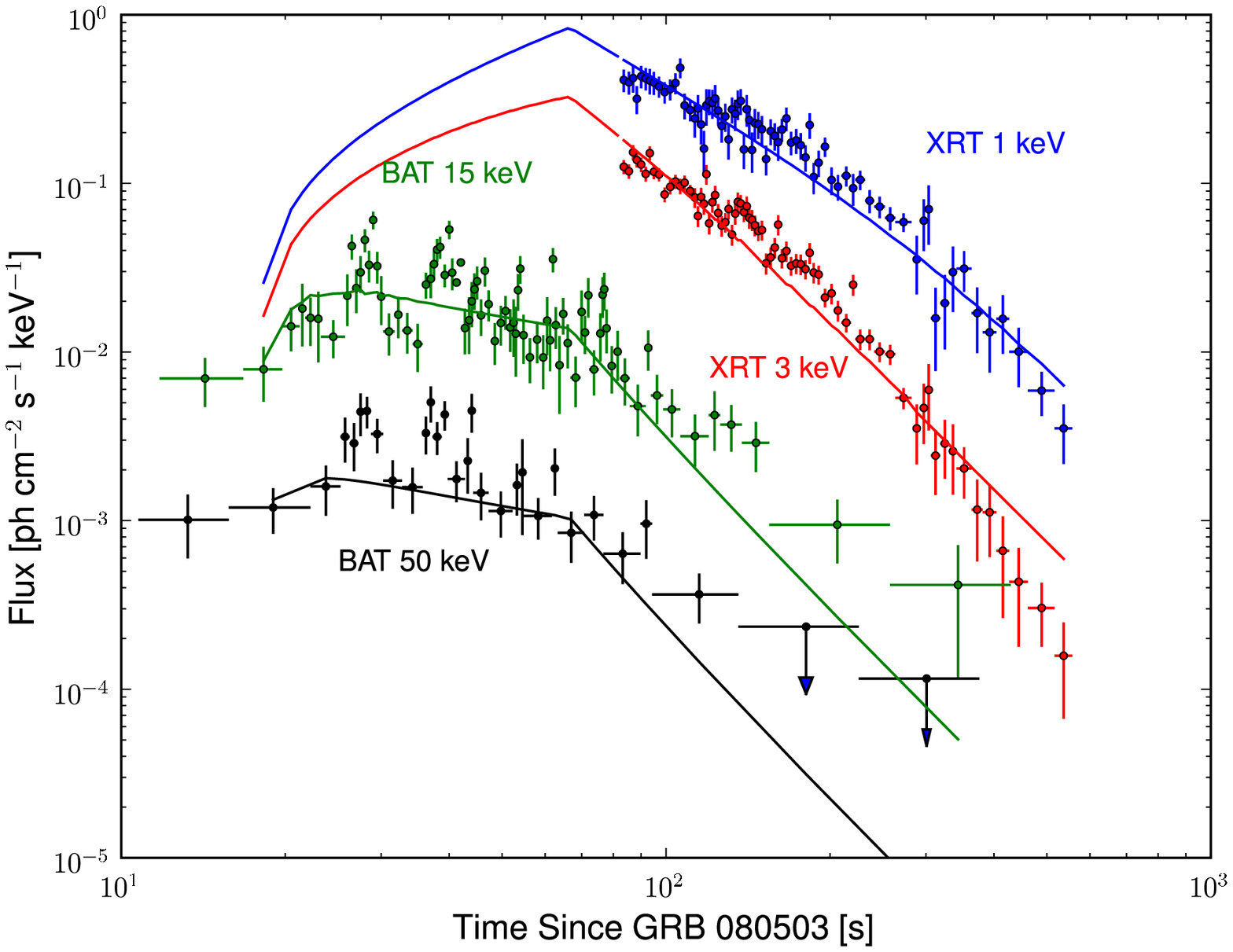}
\includegraphics[width=3.5in]{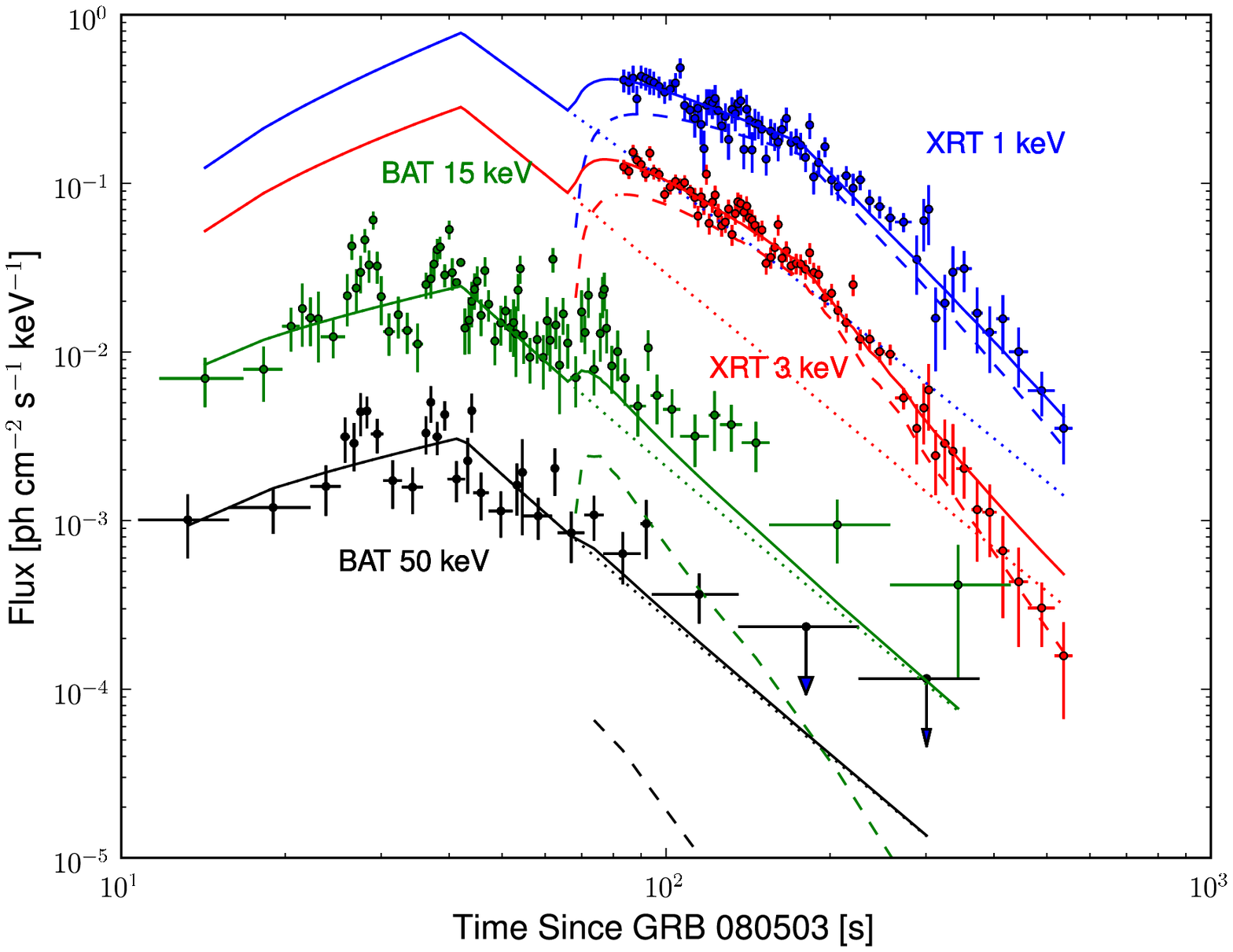}
\caption{\it One (top) and two (bottom) component model fits to the early
  GRB~080503 data from the Swift BAT and XRT.  The scaling of the counts data
  to flux units (y-axis) is approximate; however, the model fitting
  appropriately convolves the time-dependent input spectrum through the
  instrumental response matrices. In the bottom panel, dotted and dashed lines
  show the two components separately, while the solid curves show the sum of
  the two components The best-fit parameters are shown tables \ref{tab_1pulse}
  and \ref{tab_2pulses}.}
\label{fig_fits}
\end{figure}

Next, we further subdivide the BAT and XRT data each into soft and hard energy
channels: 15-50 keV and 50-150 keV for BAT, 0.3-1.3 keV and 1.3-10.0 keV for
the XRT. We rebin the data (and group the response matrices accordingly) so
that each hard and soft channel still has $S/N \ge 3$. The soft and hard
channel light curves are shown in Figure\;\ref{fig_fits}. To fit the
temporal/spectral model of GG09 to the 2-energy-channel BAT and XRT data as a
function of time, we fold the model through the response matrices at each time
bin and minimize the total $\chi^2$.  Fitting is accomplished using Markov
Chain Monte Carlo (MCMC) through the python PyMC
package\footnote{http://code.google.com/p/pymc}.

\begin{table}
\begin{center}
\caption{\it One component model: best-fit parameters}
\label{tab_1pulse}
\vspace{2mm}
\footnotesize
\begin{tabular}{rc}\hline\hline
\multicolumn{2}{c}{\it one component model} \\\hline
parameter & value\\\hline
$\beta_l$ & $-0.67 \pm 0.18$ \\
$\beta_h$ & $1.13 \pm 0.08$ \\
$t_f$ [s] &  $55.35 \pm 2.99$ \\
$t_{\rm peak}$ [s] & $66.56 \pm 3.30$ \\
$dR/R$ & $8.33 \pm 1.43$ \\
$\log{E_0}(t_{peak})$ & $0.59 \pm 0.09$ \\
$N_H$ [cm$^{-2}$] & $(8.7\pm 2.2)\times 10^{20}$  \\
$a$ & $0.67 \pm 0.17$ \\\hline
$\log{\sigma_0}$ (BAT and XRT) & $-1.27 \pm 0.08$ \\\hline
\end{tabular}
\end{center}
\end{table}

\begin{table}
\begin{center}
\caption{\it Two-component model: best-fit parameters}
\label{tab_2pulses}
\vspace{2mm}
\footnotesize
\begin{tabular}{rcc}\hline\hline
\multicolumn{3}{c}{\it two component model} \\\hline
parameter & $1^{\rm st}$ pulse  & $2^{\rm nd}$ pulse   \\\hline
$\beta_l$ & \multicolumn{2}{c}{$0$  (fixed)}   \\
$\log{norm.}$ & $-4.69\pm 1.19$ & $-0.89\pm 0.47$ \\
$\beta_h$ &  $0.63\pm 0.08$ &  $2.13\pm 0.33$   \\
$t_f$ [s] &  $37.23\pm 2.45$ &  $125.67\pm 10.37$   \\
$t_{\rm peak}$ [s] &  $42.12\pm 2.24$ &  $176.35\pm 9.11$  \\
$a$ &  $1.311\pm 0.26$ &  $-0.38\pm 0.19$  \\
$\log{E_0}$ & $1.0\pm 0.3$ & $0.3\pm 0.1$ \\
$dR/R$ & \multicolumn{2}{c}{$6.67 \pm 0.39$} \\
$N_H$ [cm$^{-2}$] & \multicolumn{2}{c}{$(2.0\pm0.4)\times 10^{21}$} \\\hline
$\log{\sigma_0}$ & \multicolumn{2}{c}{-1.63 $\pm$ 0.10} \\\hline
\end{tabular}
\end{center}
\end{table}

At the beginning of the RDP the flux is expected to be dominated by the last
pulse. However, at later times other pulses are expected to become dominant,
and the contribution of each pulse to the flux at late times ($t-t_0 \gg
\Delta t$) scales as $\sim F_{peak}T_f^{2+\beta}$ when $\beta$ is constant
among the pulses: higher (with a larger $F_{peak}$) or wider (with a larger
$T_f$) pulses will dominate. The width of the pulse having a larger power
($2+\beta \sim 4-5$ for $\beta \sim 2-3$) in the relative contribution to the
flux, it will be the most important parameter to find the dominant pulse. This
is not true if the high energy spectral slope is varying among pulses, but we
assume that such variations are small enough so that the contribution at late
times of narrow pulses are still small compared to larger ones. Given the
shape of the BAT lightcurve of GRB080503, it thus seems most natural to explain
it by one or two broad pulse(s) superimposed by narrower ones that would
account for the smaller timescale variability observed (such as the spikes at
$28\;$s and $36\;$s), but whose contribution to the RDP would be negligible:
the ratio of their width being of the order of $10/60\approx 0.16$, the ratio
of their contribution to the tail of the prompt emission would be about
$0.16^{4} \approx 7\;10^{-4}$ or less. The results presented here will
describe only the broad(s) pulse(s) fitting and results, as in it lay the most
important physics.

\subsection{Single Pulse Model}

We begin by fitting a single pulse BAT emission model to study the expected
late-time emission in the XRT bands. Initial temporal parameters (rise time
$t_f$ and peak time $t_{\rm peak}$) are chosen to crudely reproduce an
envelope of emission containing the BAT light curve. We then allow these and
the other model parameters to vary (Table \ref{tab_1pulse}).

We find that a single pulse model tends to yield a mediocre fit to the data.
The fit requires a systematic error term (in addition to the measured error)
of 30\% in order to yield $\chi^2 = \nu$. This error is not driven by the BAT
time-variability alone (although this contributes). There is clear difficulty
in fitting the smooth light curve break in the X-ray bands and also
potentially in fitting the high flux in the BAT after $t\approx
150$s. However, we note that the error bars on the late-time BAT flux are
large. The limits plotted in Figure \ref{fig_fits} (top) for the BAT hard channel
are at the $2\sigma$ level.

\begin{figure}
\begin{center}
\includegraphics[width=0.5\textwidth,height=0.5\textwidth]{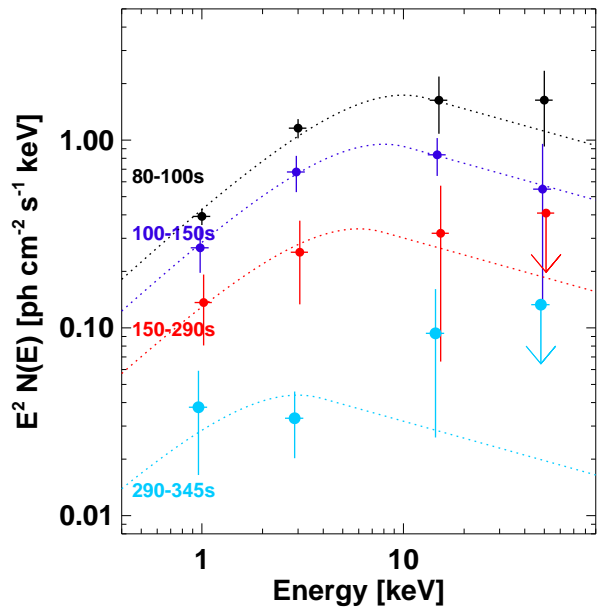}
\caption[time-resolved spectra]{\emph{Spectra of GRB080503 at different time
    interval during the RDP when both gamma-ray and X-ray are observed:
    $80-100\;$s (black), $100-150\;$s (dark blue), $150-290\;$s (red) and
    $290-345\;$s (cyan). The dotted curve show for comparison a Band function
    spectrum with $\beta_l=0$, $\beta_h=1.3$, and $F_{\nu}(E_{pk}) \propto
    E_{pk}^2$, which is roughly expected from HLE.}}
\label{fig_spectrum_evolution}
\end{center}
\end{figure}

\subsection{Double Pulse Model}

Because the single pulse model has difficulty reconstructing the slow BAT
decline as well as the rapid X-ray break, we attempt fitting a model with two
pulses. As displayed in Figure \ref{fig_fits} (bottom), the data are considerably
better fit with two pulses, the parameters for which are given in Table
\ref{tab_2pulses}. In the fitting, we tie several of the parameters from pulse
B to those from pulse A. We also fix the low energy Band model index
$\beta_l=0$, which is typical for GRBs. A large value for that index from the
single pulse model may further suggest the inapropriateness of that model.

The data are fit by a soft early pulse (peaking at $t=42$s with $E_0=1.0$ keV)
and a softer later pulse (peaking at $t=175$s with $E_0=0.3$ keV). The BAT
emission is dominated by the early pulse, whereas the late pulse provides the
temporal break observed in the XRT bands around $185\;$s, which in this case
is the beginning of the high latitude decay of the second pulse.

We have attempted fits of 3 or more pulses to the data as well. These do not
appear to significantly improve the fit, which even for the two pulse model
requires a 20\% systematic uncertainty. Apart from the possible slow BAT decay
at $t>150$s, this extra uncertainty appears to be needed to account for light
curve variability, which is not treated in the model.

\section{Discussion} \label{sec_discussion}

The main problem in fitting the data with a single pulse is the very steep
decay of the hard X-ray band combined with the relatively shallow decay of the
BAT bands. In order to find a reason for such a behaviour, we extracted
spectra at different time intervals ($80-100\;$s, $100-150\;$s, $150-290\;$s
and $290-345\;$s) along the RDP lightcurve (when data in the four BAT and XRT
bands are available) to probe the spectral evolution (see figure
\ref{fig_spectrum_evolution}, where from earliest to latest spectra the colors
are black, dark blue, red and cyan). Although the data are marginally
consistent with a Band function with constant slopes and decreasing $E_{\rm
  peak}$, there is evidence for a departure from a Band spectrum to a spectrum
with a concave shape beginning at $t \gtrsim 150$ s. Such spectral evolution
would strongly suggest that at times $\gtrsim 150\;$s a second component
contributes to the flux; this would naturally explain the distinct X-ray and
gamma-ray behaviours of the RDP, accounting for the rapid X-ray decline and
slow gamma-ray decay.

The fit with two pulses gives much better results - as expected since
two physical components seem to contribute to the flux. In this two
pulse modelling, the first part of the emission ($t \lesssim 65\;$s)
is accounted for by a first pulse, and the second part, including the
whole RDP, by the second component. The X-ray break at $\sim 185\;$s
is the start of the high latitude decay of the second pulse, and the
smooth break along the rest of the RDP is due to the sweeping up of
the peak energy of the spectrum across the observed energy bands (as
explained in \S\;\ref{sec_modeldescription}). Due to the reappearance
at late time ($t \gtrsim 300\;$s) of the first component above the
second one, the steep decay in the hard X-ray band is still not very
well accounted for. However, when looking at the 4-points spectra, one
can see that the high and low energy slopes seem to change (on top of
the appearance of a second component): from the earliest to the latest
spectrum, the low energy slope seems to decrease with time (the low
energy slope of the $\nu F_{\nu}$ spectrum becomes shallower with
time), and the high energy slope also seems to steepen. This may
explain why the introduction of a second component alone (with fixed
Band function parameters) cannot reproduce both the steep hard X-ray
and the slow gamma-ray decline, even with a very steep high energy
photon index ($\beta_h \approx 2.13$): evolution of the Band function
slopes seem required.

The slow gamma-ray decay could alternatively be due to a series of
spikes that are reasonably narrow (with $\Delta t < t$), unresolved (due
to the wide late time BAT time bins) and hard (dominating at soft
gamma-rays but with a small X-ray flux), so that we see only their smooth
envelope, while their tail would hardly contribute to the observed X-ray
decay. We do not directly model this, however, since it would introduce
far too many new free parameter, and thus not provide a very stringent
test for such a model.

\begin{table}
\begin{center}
\caption{\it Dependence of the parameters $a$ and $d$ (with $L'_{\nu'_p}
  \propto R^{a}$ and $\nu'_p \propto R^{d}$) on $u$ (with the strength of the
  shocks in the outflow being parametrized by the relative uptstream to
  downstream four-velocity as $U_{rm ud} \propto R^u$), and values of $u$ and
  $d$ for the fitted value $a_2=-0.38\pm 0.19$ of the second pulse of the two pulses model.}
\label{tab_a_d_udep}
\vspace{2mm}
\footnotesize
\begin{tabular}{rcc}\hline\hline
\multicolumn{3}{c}{\it Equipartiton field}                \\\hline
             & $u \ll 1$       & $u \gg 1$                \\\hline
$a(u)$       & $1-2u$          & $1-2u$                   \\\hline
$d(u)$       & $5u-1$          & $3u-1$                   \\\hline
$u(a_2)$     & $0.69\pm 0.095$ & $0.69\pm 0.095$          \\\hline
$d[u(a_2)]$  & $2.45\pm 0.475$ & $1.07\pm 0.285$          \\\hline
\hline
\multicolumn{3}{c}{\it Advected field}                    \\\hline
             & $u \ll 1$      & $u \gg 1$                 \\\hline
$a(u)$       & $1-u$          & $1-2u$                    \\\hline
$d(u)$       & $4u-1$         & $3u-1$                    \\\hline
$u(a_2)$     & $1.38\pm 0.19$ & $0.69\pm 0.095$           \\\hline
$d[u(a_2)]$  & $4.52\pm 0.76$ & $1.07\pm 0.285$           \\\hline
\end{tabular}
\end{center}
\end{table}

We had kept $a$ as a free parameter since it did not complicate the modeling
and could test for deviations from its simplest form. Indeed, none of the
values obtained for the fits are perfectly consistent with $1$. Considering
only the two pulse fit (the one pulse fit not being good enough to have any
physical significance), the values of $a$ for the first and second pulses
(pulse number indicated by a subscript number) are $a_1=1.31\pm 0.26$ and
$a_2=-0.38\pm 0.19$. Since the value $a=1$ was obtained under the assumptions
of electrons cooling fast by synchrotron emission in a coasting outflow, this
means that at least one of the above assumptions is not true in the case of
GRB080503. The most natural assumption to relax is that the strength of the
shocks in the outflow is constant. Assuming that it varies with radius, namely
parametrizing the relative uptstream to downstream four-velocity as $U_{\rm
  ud} \propto R^u$ (while the Lorentz factor of the emitting shocked region
remains constant), then $a = a(u)$ and $d = d(u)$ where $L'_{\nu'_p} \propto
R^a$ and $\nu'_p \propto R^d$. The dependence of $a$ and $d$ on $u$ are
summarized in table \ref{tab_a_d_udep}, where we have considered two options
for the magnetic field in the emitting region: (i) a field strongly amplified
at the shock that holds a constant fraction $\epsilon_B$ of the internal
energy (equipartition), and (ii) a pre-existing magnetic field advected from
the central source (scaling as $R^{-1}$ upstream) and merely compressed at the
shock (advected field). The values of the parameters $u$ and $d$ corresponding
to the value $a_2=-0.38\pm 0.19$ of the second pulse are given as well (since
the value of $a$ for the first pulse is close to its fiducial value
$a=1$). One can note that the value of $a$ inferred form the fit implies a
value of $d >1$, meaning that $\nu'_p$ rises with radius at least linearly,
which is very different from the basic model where $\nu'_p \propto R^{-1}$. We
stress, however, that the parameters $a$ and $d$ affect mainly the rising
parts of the pulses and hardly affect the decaying parts. Since we do not have
strong constraints on the time variation of the peak frequency during the
rising parts of the pulses, parameter values similar to the ones we infer may
be able to reasonably fit the data.

\section{Conclusion}

We have explored the extended emission and very long RDP of GRB080503 in the
light of the realistic physically motivated model for the prompt and latitude
emission from \citet{GG09}. Neglecting narrow pulses that do not affect the
high latitude emission, we fitted the broad underlying envelope of the extended
emission and its RDP, expecting a simple HLE behavior, motivated by the
results of \citet{peretal09}. The first attempt, using a single pulse,
failed to account for both the slow gamma-ray and very rapid hard X-ray
declines. A close look at the spectral evolution during the RDP (at times
$\gtrsim 150\;$s) showed evidence of a strong spectral evolution: the emergence of a
second spectral component dominating the high energies is spotted, as well as
evolution of the spectral slopes of the (assumed) Band function spectrum. This
led us to fit the data with two pulses, the first one accounting for the
gamma-ray envelope and decay phase, the second one accounting for the X-ray
decay phase. The fit thus obtained is far better - even if not perfect, since
we are still not accounting for the smaller variability. Adding a third
pulse does not improve the fit, which suggests two pulses are enough to
account for the data (as suggested by the spectral evolution). However, the
fit we obtained is still far from perfect (the hard X-ray decay is still too
steep at the latest times), the model being unable to account for the
evolution of spectral slopes, and the fitting results showing discrepancies
with its simplest form. This points out that the simplest form of GG09 model
for HLE is not detailed enough to account for the complex RDP of
GRB\,080503. The mechanism at work here may thus still be HLE, but since this
model has some difficulties and requires many degrees of freedom in order to
produce a reasonable fit, this may suggest that another mechanism might be at
the origin of the RDP in GRB\,080503.

A previous attempt at fitting GG09 model to a short pulse with
extended emission was done on GRB050724 by \citet{willetal09}. They
also could not obtain a good fit of the extended emission in both BAT
and XRT band, the model not being able to account for a slow BAT decay
and a steep XRT decay during the RDP - similarly to what is observed
with GRB080503. Such feature may thus be a characteristic of RDP from
extended emission, but more bursts should be studied before any
conclusion can be drawn. \citet{willetal09} argue that HLE is a valid
explanation for the RDP of GRBs without extended emission. If there is
indeed a difference in the RDP of bursts with and without extended
emission, then the properties of extended emission would be different
from the prompt emission, which might suggest they have a different
physical origin. GRB080503 may thus be the first stone paving the way
to a better understanding of extended emissions in Gamma-ray bursts by
studying their RDP. However, one should keep in mind that, on top of
having been studied more carefully than other bursts, the RDP of
GRB080503 is uncommonly long, meaning that maybe the range on which
rapid decay is usually observed is not large enough to show deviations
from the high latitude decay. In this case this should be taken as a
caveat for drawing conclusions from model fitting to the rapid decay
phase of GRBs.

\section*{Aknowledgement}

NRB is supported through the GLAST Fellowship Program (NASA Cooperative Agreement: NNG06DO90A).
J.G. gratefully acknowledges a Royal Society Wolfson Research Merit Award.


\label{lastpage}

\end{document}